# Isovalent substitution effects of arsenic on structural and electrical properties of iron-based superconductor $NdFeAsO_{0.8}F_{0.2}$

Z. Alborzi and V. Daadmehr*

Magnet and Superconducting Res. Lab., Faculty of Physics & Chemistry, Alzahra University, Tehran 19938, IRAN

Zahra_alborzi@yahoo.ca, *daadmehr@alzahra.ac.ir

# Abstract

In this paper, nominal compositions of $NdFeAsO_{0.8}F_{0.2}$, $NdFeAs_{0.95}Sb_{0.05}O_{0.8}F_{0.2}$ and $NdFeAs_{0.95}P_{0.05}O_{0.8}F_{0.2}$ were prepared by one-step solid-state reaction method. The structural, electrical and morphological properties of samples were characterized through the XRD pattern, the 4-probe method and SEM, respectively. The crystal structure of our samples was tetragonal with P4/nmm:2 symmetry group. Also, the (x, y, z), occupancy of ions and lattice parameters were changed by isovalent substitution of Phosphorus (P) and Antimony (Sb)in the $NdFeAsO_{0.8}F_{0.2}$ sample. The α, β bond angles and Fe-As bond length are changed from the corresponding value of α and β regular FeAs4-tetrahedron by isovalent doping P/As and Sb/As, that they are effective on the superconductivity transition temperature. The microstrain and crystalline size of samples were studied by the Williamson Hall method. The superconducting critical temperatures were attained at 56 K and 46 K for $NdFeAsO_{0.8}F_{0.2}$, $NdFeAs_{0.95}Sb_{0.05}O_{0.8}F_{0.2}$, respectively. The $NdFeAs_{0.95}P_{0.05}O_{0.8}F_{0.2}$ showed the structural transition temperature at 140 K. It seems that there is a relation between the superconductivity and shrinkage of the crystal lattice. The flake-type of grains was observed by SEM pictures of samples.

## 1. Introduction

For the first time in 2006, superconducting transition temperature ($T_C$) was observed in a new category of superconductors called iron-based superconductors (IBs), in pure and fluorine-doped LaOFeP at $T_C$=3.2 K[1]. After, the report of superconductivity at 26K in fluorine-doped LaFeAsO was registered as high transition superconductors in IBs [2]. Because of the existence of vortices and high superconducting transition temperature in IBs, they are in unconventional superconductor and high temperature superconductors, respectively[3].

IBs have some of the most important and unique characteristics such as: parent material: antiferromagnetic metal, impurity robust $T_C$, multi-band nature of Fe-3d, small anisotropy, large upper critical field and advantageous grain boundary nature[4].

IBs are divided in to four subgroups, i: $FeSe_{1-x}Te_x$ (11-type)[5], ii: LiFeAs (111-type)[6], iii: $MFe_2As_2$ (M=Ba, Sr, K. 122-type)[7,8], iv: ReFeAsO (Re=La, Nd, Sm. 1111-type)[9-11]of course two other subgroups were introduced for this category of superconductors: $MFeAs_2$ (M=Ca, La. 112-type)[12] and $A_2Fe_4Se_5$ (A = K, Rb, Cs. 245-type)[13]. In these family, the subgroup of 1111-type was the first discovered among all IBs [1], so it is also the most studied family with the biggest variety of compositions.

Also, it should be noted that there is the highest transition temperature of superconductivity in this group of IBs [14-17], it can be due to the interlayer spacing of FeAs layers and the bond angle of Pn(Ch)–Fe–Pn(Ch) in this structure[18,19]. The substitution effect on this compound has showed the change in transition temperature[20-23]. In IBs, there are electron[24] and hole doping[25,26]. A particular characteristic of doping into IBs is isovalent doping. There are two models for doping. Partial substitution in superconductivity planes of FeAs instead of Fe[27] or As[28], and another is in Re (O, F) Layer as charge carrier planes[29]. For isovalent doping, atoms are implemented into the material without change of the charge carrier concentration.

The effects of chemical pressures and bond covalency due to isovalent substitution of P/As in LaFeAs1-xPxO were investigated by C. Wang et al. [30]. In ref. [35], superconductivity emerges in the region of 0.2<x<0.4 with the maximum transition temperature $T_C$ =10.8 K for x=0.25. In 2010, C. Wang et al.[31] studied the co-doping of Sb and F on the structural and superconducting properties of $LaFeAs_{1−x}Sb_xO_{1−y}F_y$. They result that the Sb doping hardly influences the spin density wave (SDW) anomaly. In 2014, Q. Ji et al.[26] reported the enhancement of superconductivity transition temperature by Sb doping in iron pnictide superconductor $Pr_{1-x}Sr_xFeAsO$. They obtained that the lattice constants remain almost unchanged and contracted when the Sb and P doped in this structure. C. Wang et al.[30] investigated the effect of chemical pressures and bond covalency in $LaFeAs_{1-x}P_xO$ by the isovalent substitution of P/As. In this work, superconductivity emerges in the region of 0.2<x<0.4 with the maximum $T_C$ =10.8 K for x=0.25. In 2010, C. Wang et al.[31] studied the co-doping of Sb and F on the structural and superconducting properties of $LaFeAs_{1−x}Sb_xO_{1−y}F_y$. They result that the Sb doping hardly influences the SDW anomaly. In 2014, Q. Ji et al.[26] reported the enhancement of superconductivity by Sb-doping in the hole-doped iron-pnictide superconductor $Pr_{1-x}Sr_xFeAsO$. They obtained that the lattice constants remain almost unchanged and contracted when the Sb and P doped in this structure.

In 2008, Ren et al. reported a new quaternary iron-arsenide superconductor $NdFeAsO_{1−x}F_x$, with the onset resistivity transition at 51.9K [15]. One of the major goals of the researchers is to achieve high temperature superconductivity in these superconductors. To our knowledge the highest Tc transition in the $NdFeAsO_{1−x}F_x$ material is obtained and reported at 55 K by controlling of As atmosphere in ambient pressure [32], at 54 K by oxygen-deficiency in fluorine-free of $NdFeAsO_{1-y}$ [16] and at 55.1 K by doping of chemical elements [29] so far. It can be concluded that the construction method and the doping of chemical elements are more effective in achieving higher transition temperatures. Motivated by this line of reasoning, we have tried to improve the synthesis method and to get a higher superconductivity transition temperature. For this purpose, we have synthesized a series of $NdFeAsO_{0.8}F_{0.2}$ samples using solid state reaction technique. In previous our work, the construction method improved by thermogravimetrically analysis TGA and we were able to record the highest transition superconductivity temperature for this nominal composition $NdFeAsO_{0.8}F_{0.2}$ at 56 K[14].

In the following, our goal is to focus on the effect of the isovalent substitution elements on Arsenic in this nominal composition. For this purpose, the influence of isovalent substitution of Phosphorus (P) and Antimony (Sb) for Arsenic (As) on the structural and electrical properties of nominal composition of $NdFeAs_{1-x}M_{x}O_{0.8}F_{0.2}$ (M=P, Sb) is investigated by using X-ray powder diffraction for structural properties along with the 4-probe method for electrical measurement. We study the XRD patterns for calculation of crystal structure parameters such as occupancy, bond length, bond angle, microstrain and crystalline size which is determined by Rietveld refinement by materials analysis using diffraction (MAUD) software[33]. The surface morphology of grains will be studied by Scanning Electron Microscopy (SEM).

## 2. Materials and method

Samples with nominal compositions of $NdFeAs_{1-x}M_xO_{0.8}F_{0.2}$ with x=0.0 ,0.05 for M=Sb or P, labeled as Nd-1111, Nd-Sb0.05 and Nd-P0.05 respectively, were synthesized by one step solid-state reaction method. At first, the stoichiometric amounts of Nd (99.99%), As (99.99%), $Fe_2O_3$ (99.9%), Fe (99.9%), $FeF_3$ and red phosphorus P (99.99%) or antimony Sb (99.5%) powders were mixed according on the equation (1):

$$Nd + (1-x)As_{1-x} + xM + \frac{6}{15}Fe + \frac{4}{15}Fe_2O_3 + \frac{1}{15}FeF_3 \rightarrow NdFeAs_{1-x}M_xO_{0.8}F_{0.2} \quad (1)$$

The mixed powder was pressed to 45 kg/cm$^2$ and further encapsulated in an evacuated quartz tube. All stages of procedure were performed in glove box in the nitrogen atmosphere. Then, the sealed and evacuated quartz tube was heated with a slow heating rate through 4 stages: (i) 5 h at 350 °C, (ii) 14 h at 640 °C, (iii) 20 h at 880 °C, and (iv) 20 h at 1150 °C. After the heat treatment, the sample comes in powder form in the quartz tube. Again, the powder was pressed and was encapsulated in an evacuated quartz tube. The heat treatment was repeated. Then furnace was cooled slowly down to room temperature. After this stage, the obtained compound was hard and in black color. In the structure of $NdFeAsO_{0.8}F_{0.2}$, formation of the NdO and FeAs layers is very important, because of this, we used thermogravimetric analysis of NdAs to determine the correct thermal process[14]. Due to the thermogravimetric analysis of NdAs and negligible effect of amount of phosphorus and antimony in thermogravimetric analysis of this compound, the heat treatment was repeated again. After two stages of heat treatment, we got the black pill for these two compounds.

## 3. Results and discussion

X-Ray crystallography is a technique used for determining the crystal structure of sample. X-ray powder diffraction patterns were measured using a X'Pert PRO MPD, PANalytical Company (Netherlands) , for Cu-Kα (λ= 0.15406 nm) with a step size of 0.0260°.

The XRD patterns of the Nd-1111, Nd-Sb0.05 and Nd-P0.05 have been showed in **Fig. 1**. This type of IBs has two-dimensional layer of Fe (Pn/Ch) tetrahedra, and there are two angles in tetrahedra that they indicated by α and β (See **Fig. 3**). The β angle and its bond length of the Fe (Pn/Ch) in the tetrahedra are related to superconductivity transition temperature $T_C$[16,24,34,35].

The inset of **Fig.1** show the enlarged view of XRD patterns of the main peak (102) of the Nd-P0.05 and Nd-Sb0.05 samples keeping Nd-1111 as a reference pattern. It is observed that the main peak of P and Sb doped samples shift towards the right. In order to understand the effect of P and Sb substitution on the crystal lattice of Nd-1111 system, the lattice parameters of all the samples were calculated using the Rietveld refinement of XRD data by using MAUD software.

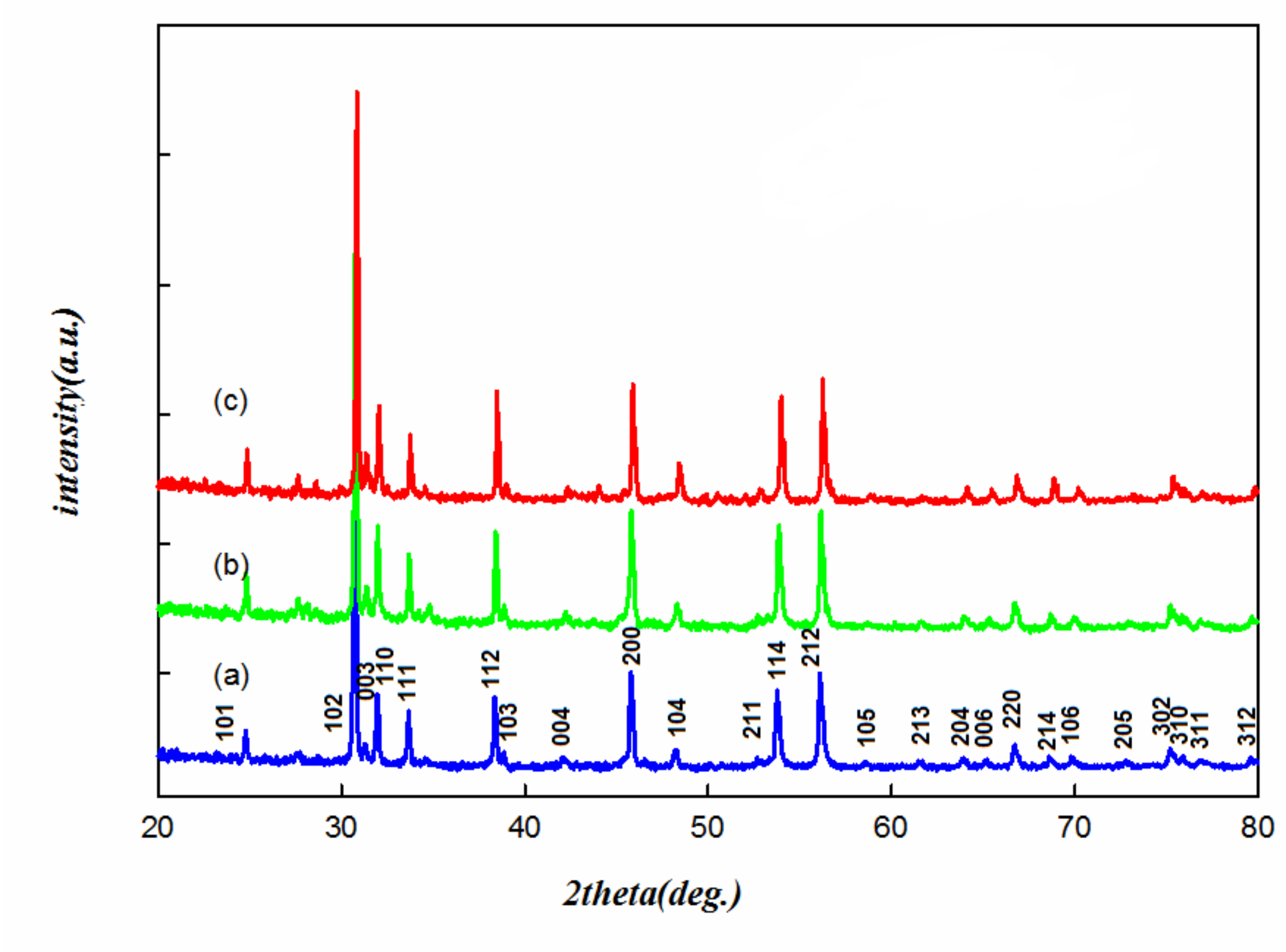

*Figure 1. XRD patterns of samples: (a) Nd-1111(blue line), (b) Nd-P0.05(green line) and (c) Nd-Sb0.05(red line)*

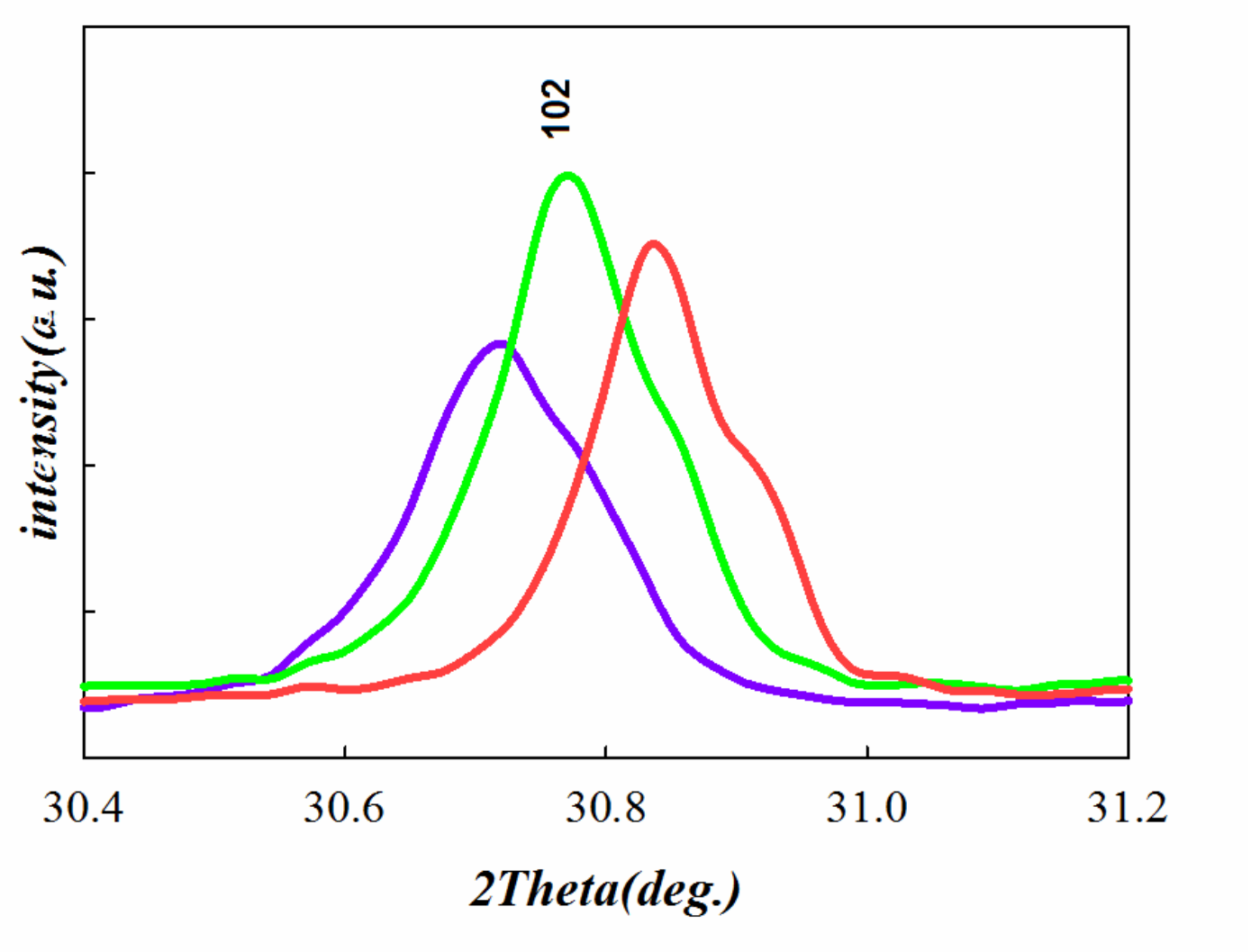

*Inset of Figure 1. Enlarged view of the main peak (102) of Nd-1111(blue line), Nd-P0.05(green line) and Nd-Sb0.05(red line)*

Rietveld analysis of the X-ray diffraction data of the Nd-1111, Nd-Sb0.05 and Nd-P0.05 samples show that the structure of samples is tetragonal with space group P4/nmm:2.

The S parameter ($S=R_{wp}/R_{exp}$) describe the fit quality, that $R_{wp}$ and $R_{exp}$ are the residual error and the expected error, respectively. Since the antimony atom can react with two oxidation states of +3 and +5, we consider all cases in Rietveld analysis for Nd-Sb0.05. Samples with nominal compositions of $NdFeAs_{0.95}Sb_{0.05}O_{0.8}F_{0.2}$ for Sb with +3 and +5 oxidation states, labeled as Nd-Sb(+3)0.05, Nd-Sb(+5)0.05, respectively. All MAUD analyzed patterns of these samples have been showed in **Figs. 2-(a-e)**. Since we want to investigate the isovalent substitution effect of the Sb/As and P/As on the Nd-1111 structure, we consider the occupancy number of As, P and Sb is float (i.e. the isovalent substituted atoms) in all Rietveld analysis. The Rietveld results of before- and after- refinement are listed in **table 1**.

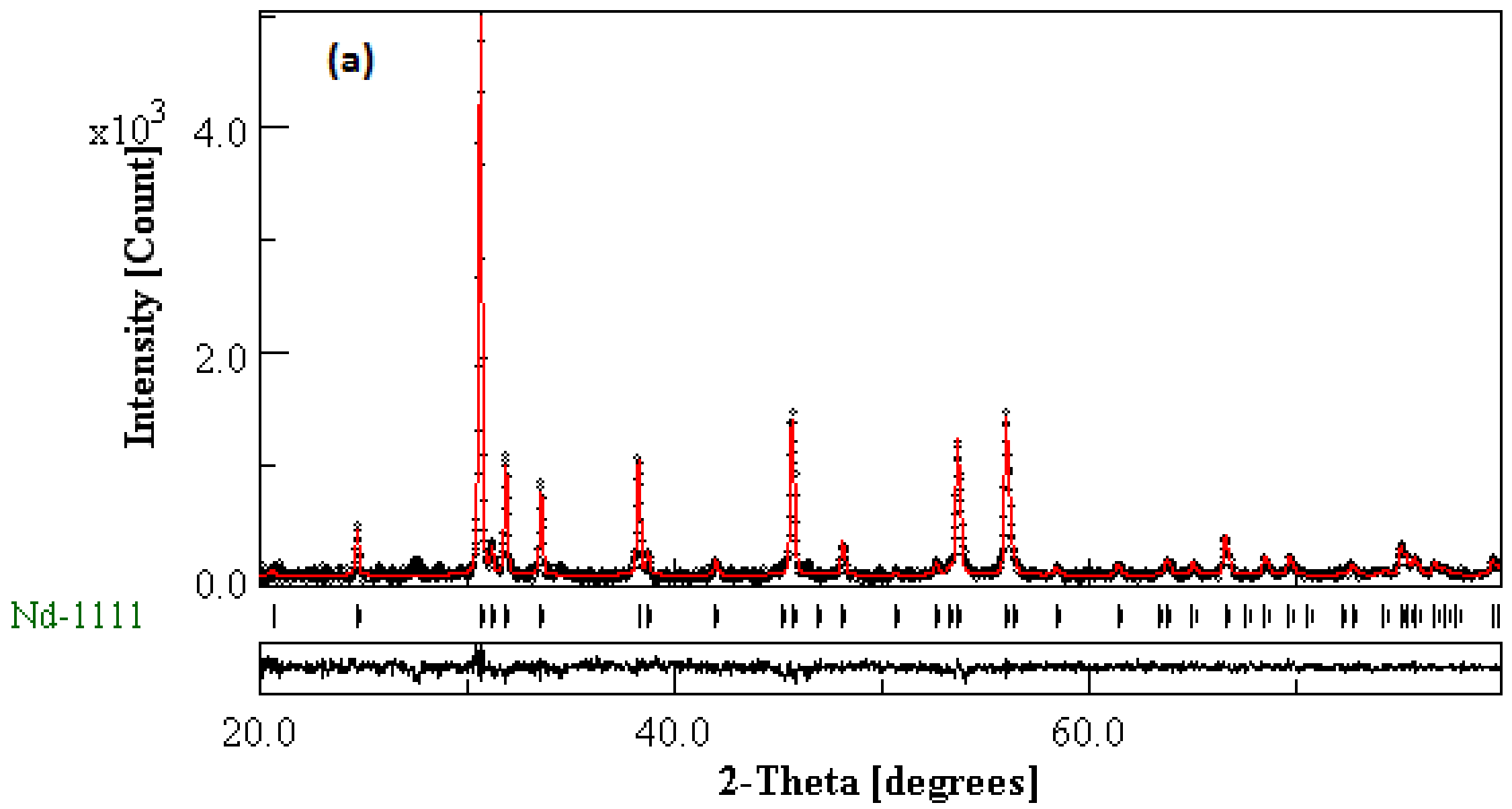

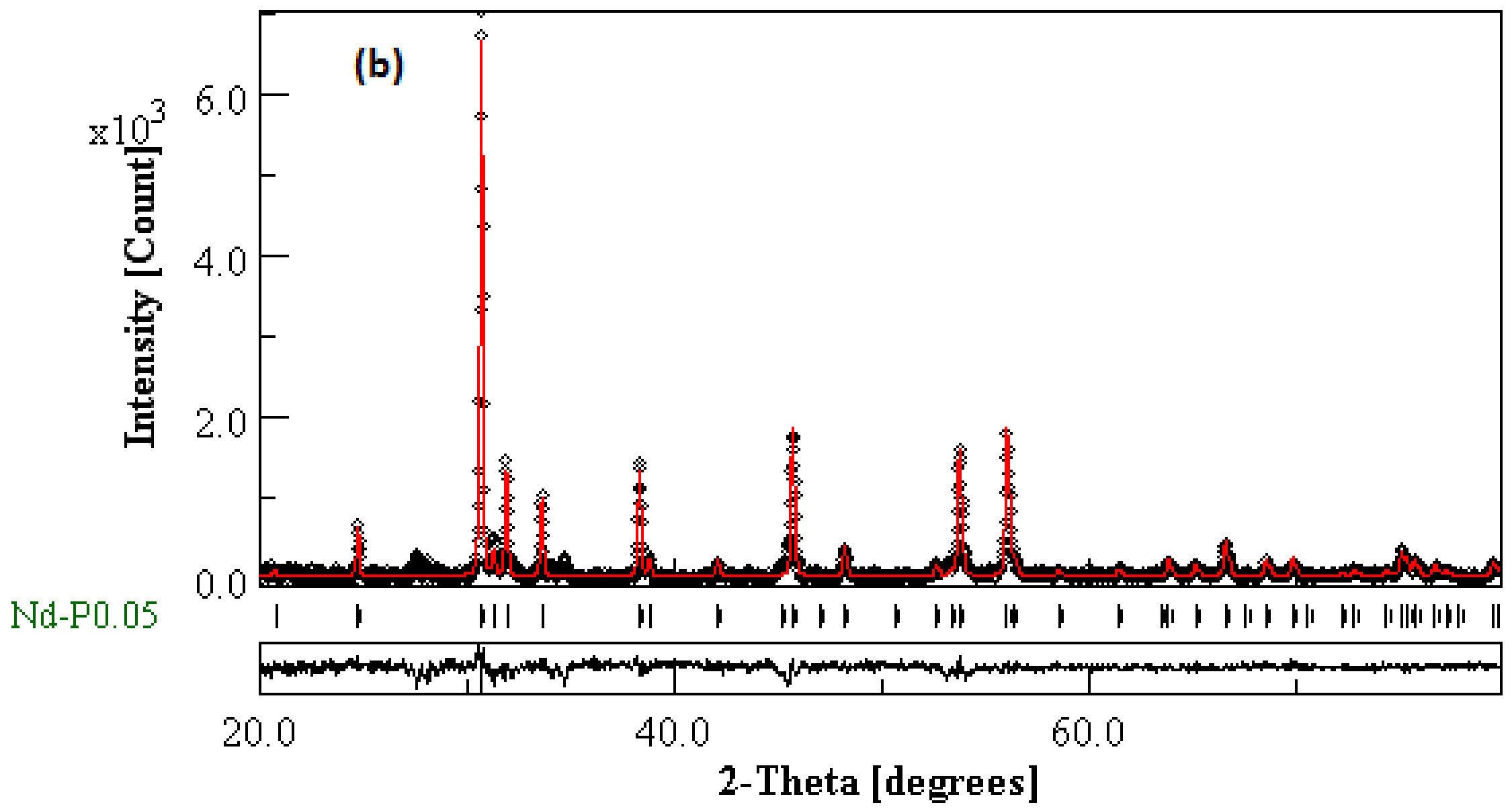

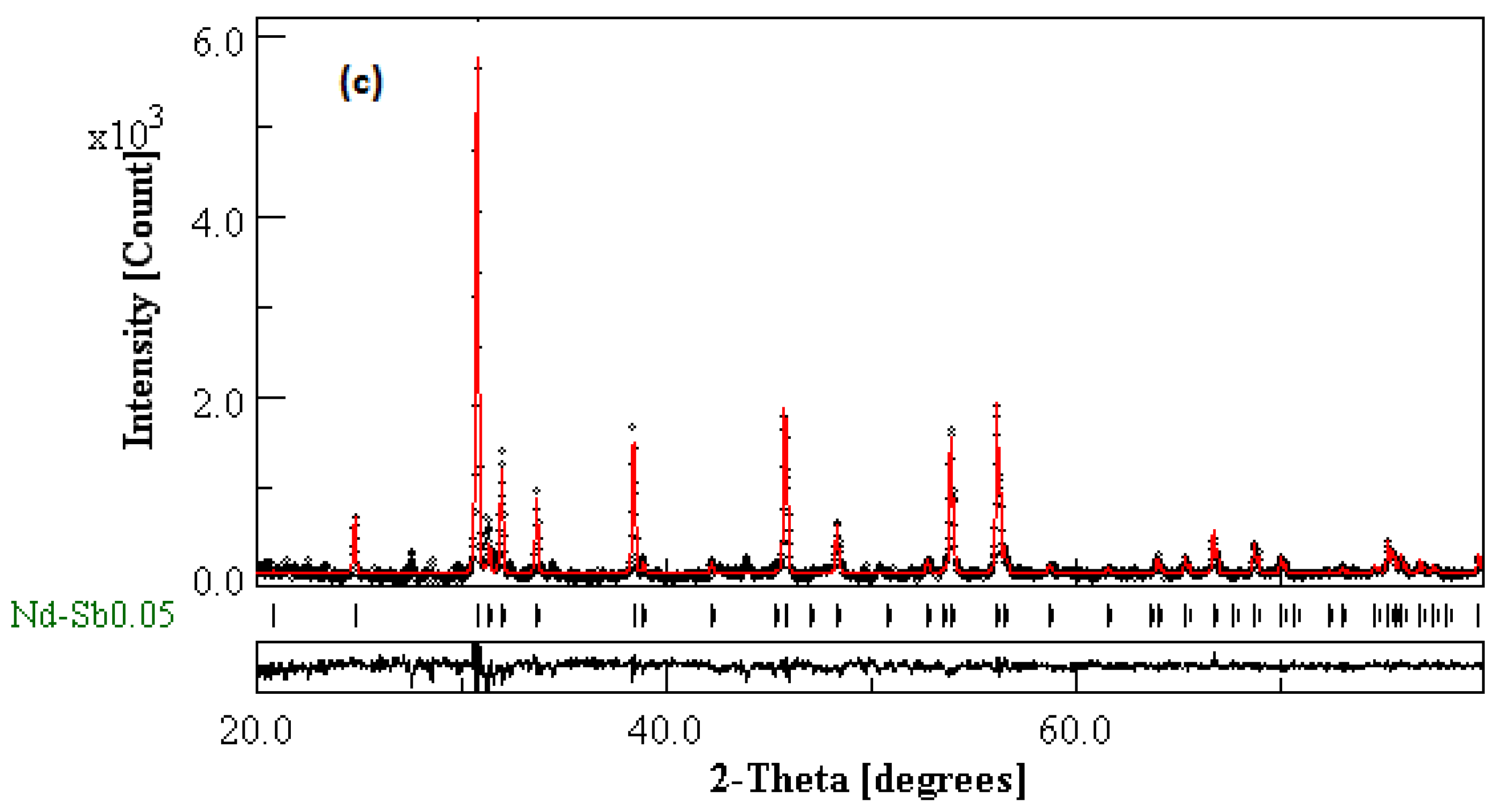

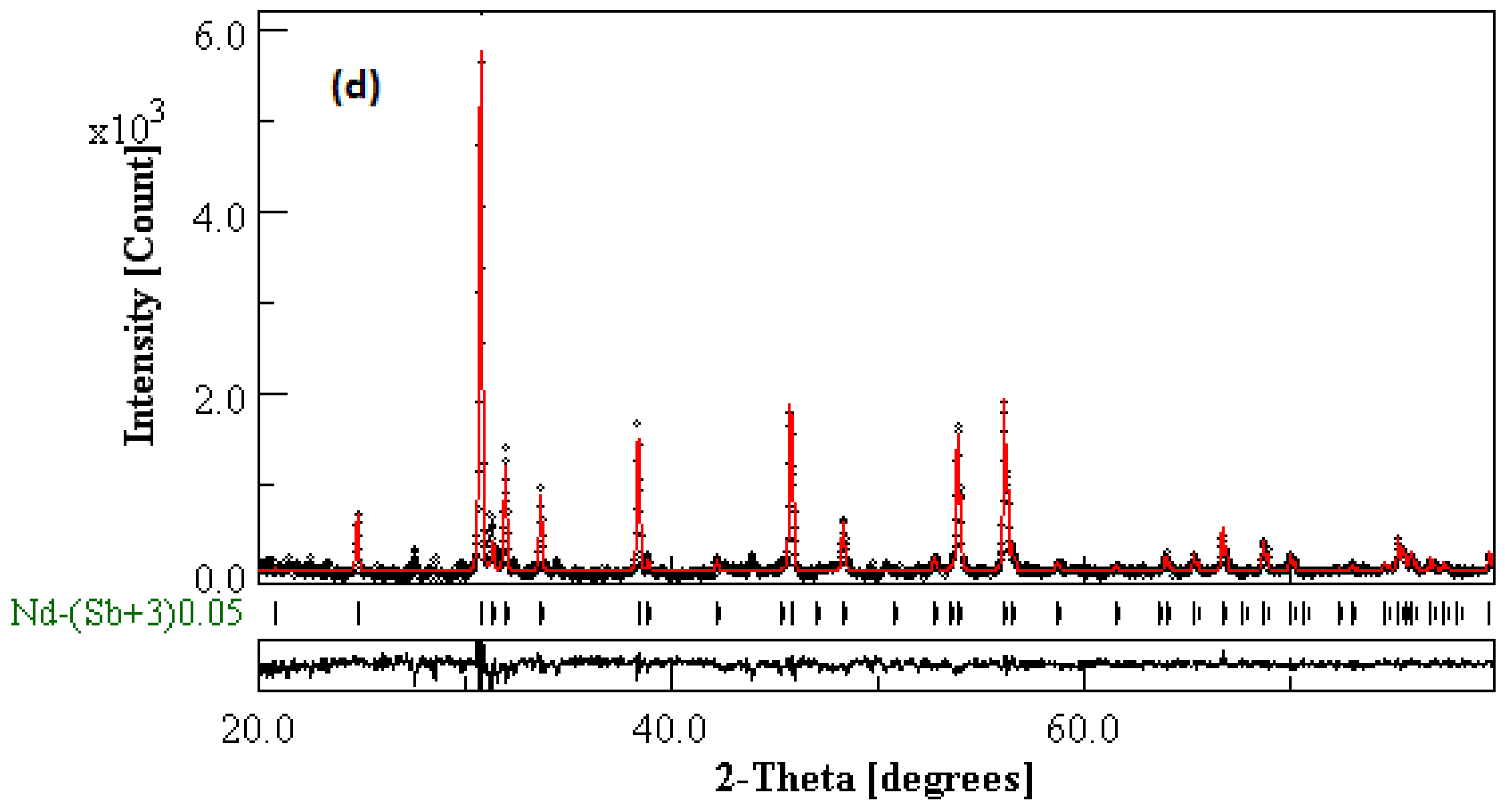

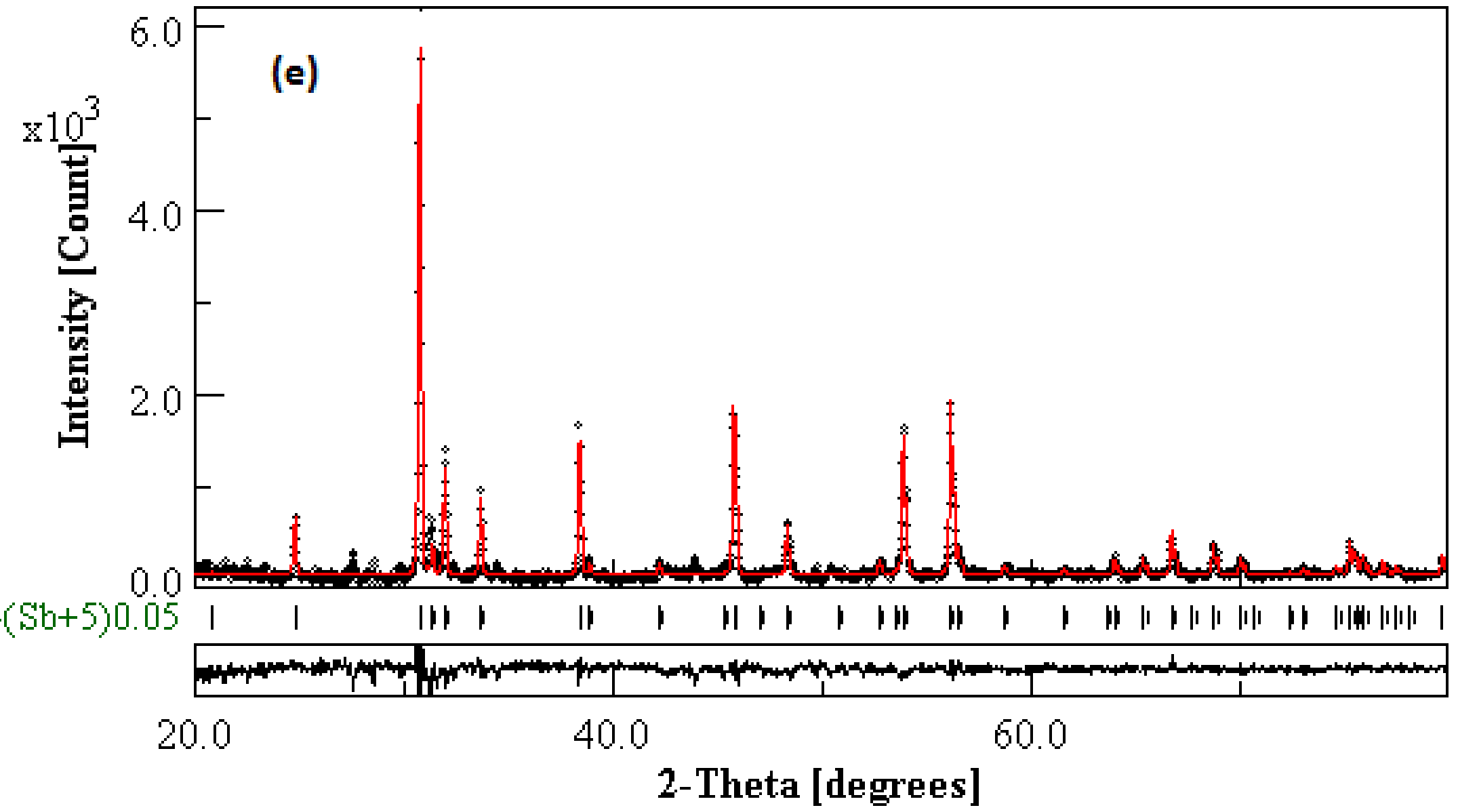

*Figure 2- X-ray powder diffraction pattern and best-fit refinement profile of (a)Nd-1111, (b) Nd-P0.05, (c) Nd-Sb0.05, (d) Nd-Sb0.05 for $Sb^{+3}$, (e) Nd-Sb0.05 for $Sb^{+5}$ observed at room temperature. The calculated pattern is shown by a red line passing though black data points in the upper portion. The short vertical black bars below the pattern indicate the positions of the allowed reflections. The black line in the lower portion shows the differences between the observed and calculated patterns.*

In MAUD software, S parameter has good fitness quality for all samples (See **table 1**). The existence of the (101), (102), (003), (110), (111), (112), (103), (004), (200), (104), (211), (114), (212), (213), (204), (006), (220), (214), (106), (205), (302), (310), (311) and (312) major Bragg peaks in the XRD patterns confirms the formation of tetragonal structure with the p4/nmm:2 space group in the Nd-1111 sample. It indicating that almost any impurity phase is in this sample.

Based on the Rietveld analysis of X-ray diffraction data of Nd-Sb0.05 and Nd-P0.05, the crystal structure of the samples was obtained as tetragonal structure with the p4/nmm:2 space group. In Nd-P0.05 and Nd-Sb0.05 samples, it can be seen two weak peaks in 27.52° and 28.59°. These mentioned peaks have not been considered by other research works[36,21,37,38]. We were refined X-ray diffraction data of substituted samples by the NdAs, FeAs, $Nd_2O_3$, NdOF, $FeAs_2$ and many undesirable phases of crystal information files (cif file) but the mentioned peaks were not covered by MAUD refinement of the Nd-P0.05 and Nd-Sb0.05 XRD patterns. According to our perfect calculations, it seems that these peaks can be covered by cif file of $NdF_2$ compound. $NdF_2$ has a cubic crystal structure with the Fm-3 space group. The (111) peak of $NdF_2$ matched with the major Bragg peak at 27.52° in our samples. Similarly, the Bragg peak at 28.59° of Nd-Sb0.05 and Nd-P0.05 samples belongs to Sb and P with rhombohedral structure with R-3m space group in pure form.

The crystal structure parameters including exact atoms position and occupancy number in unit cell for our sample were refined by MAUD software (See **table 1**).

The occupancy of As, P and Sb for after- refinement, have less amount than the considered value, since these materials are evaporated during the synthesis. In many research works, in order to compensate for the loss of As element during the synthesis, the precursors As has 5%-15% more than stoichiometric value of it [39,40].

**Table 1** presents the crystal structure parameters of unit cell for three samples. i) for N-d1111: a= 0.3967 nm and c=0.8596 nm. Substitution of P/As and Sb/As for content x=0.05 has been affected on crystal structure parameters. ii) for Nd-P0.05: a= 0.3968 nm, c=0.8582 nm and iii) for Nd-Sb0.05 a=0.3962 nm, c=0.8560 nm. **Fig. 3** presents the schematic picture of unit cell in tetragonal structure with space group of p4/nmm:2. Rietveld analysis represent the $"a"_{Nd-P0.05} > "a"_{Nd-1111} > "a"_{Nd-Sb0.05}$ and $c_{Nd-1111} > c_{Nd-Sb0.05} > c_{Nd-P0.05}$. It can be anticipated that with the increase of crystal structure parameter c of the unit cell, the superconductivity transition temperature increases based on the results of other researchers for Nd-1111 structure [41]. Decreasing of the "c" parameters can be related to the electronegativity[42] of isovalent atoms. Based on our results in table 1, in comparison of Nd-Sb0.05 with pure sample of Nd-1111, we have $Z_{Sb}(0.6563c) < Z_{As}(0.6580c)$. Moreover, the electronegativity of Sb ion (2.05) is smaller than to the electronegativity of As ion (2.18) causes the bond length of Fe-As/Sb is less than Fe-As and then α angle increases and β angle decreases. The β angle for our samples versus the β angle for regular FeAs4-tetrahedron are:

$$\beta_{\text{regular FeAs4-tetrahedron}}(109.47°) > \beta_{\text{Nd-1111}}(108.61°) > \beta_{\text{Nd-Sb0.05}}(108.26°) > \beta_{\text{Nd-P0.05}}(107.3°)$$

The effect of structural parameters on superconductivity in 1111-type IBs were investigated by C. H. Lee et al.[43]. They claimed that Superconducting transition temperatures seem to attain maximum values for regular FeAs4-tetrahedrons. Then, it can be expected that the superconductivity transition temperature for our pure sample greater than the Sb-doped sample.

In cuprates, the highest superconductivity transition temperature is observed for flat and square $CuO_2$ planes in the tetragonal structure and it is decreased by the structural distortions of the $CuO_2$ planes in the orthorhombic structure[44-47]. The 1111-type of IBs have layered structure such as cuprates and also the effect of tetrahedral distortion has been studied on superconducting transition temperature. In this type of IBs, as much as the As-Fe-As tetrahedral angle is closer to the regular tetrahedral angle (109.47°), the higher transition temperature expects for sample. Moreover, the DOS of superconducting layers in IBs in corresponds to $CuO_2$ plane in cuprates superconductor is maximum for these angles[48-51]. These results indicate relevant approach of superconducting layers for IBs and cuprate superconductors.

About the replacement of major Bragg peak (102) based on the refinement of XRD data all diffraction peaks clearly shift to higher angles with increasing of “a” unit cell parameter. Our result confirmed the consistency of dopant value of isovalent atom with their occupancy number and the replacement of origin peaks in XRD pattern of samples.

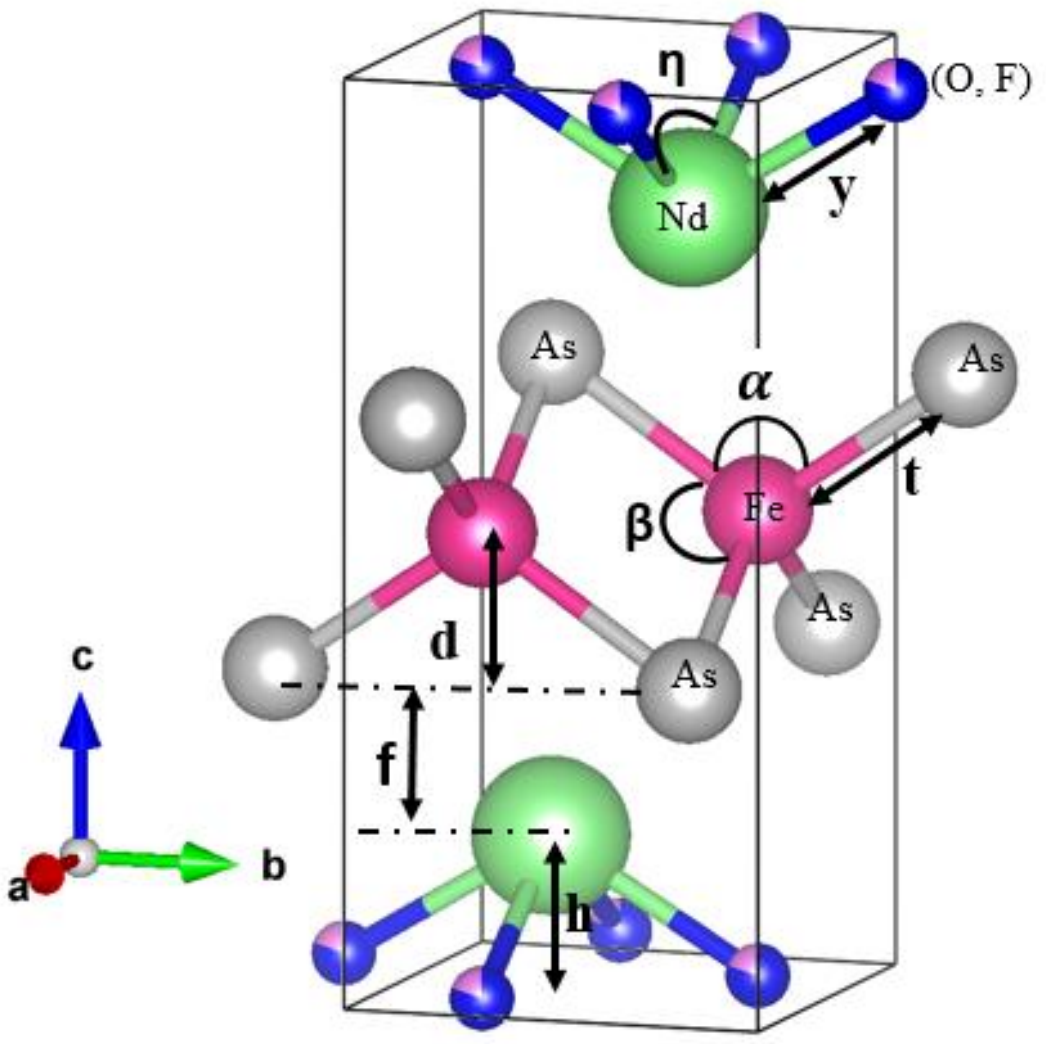

*Figure 3. The schematic picture of NdFeAsO$_{0.8}$F$_{0.2}$unit cell such as bond angles α: As-Fe-As (deg.), β: As-Fe-As (deg.), η: (O, F)- Nd- (O, F) (deg.) and Bond lengths t: Fe-As (nm), y: Nd-O (nm), 2d: thickness of Fe-As layer (nm), 2h: thickness of Nd-O layer(nm), F: distance of Fe-As and Nd-O layers(nm)*

Table 1. Refined crystal structure parameters of *Nd-1111, Nd-P0.05, Nd-Sb0.05, Nd-Sb(+3)0.05 and Nd-Sb(+5)0.05* by MAUD analysis in before- and after- refinements.

**Sample: Nd-1111**

| a(nm) | c(nm) | V (nm$^3$) | Space group |
|---|---|---|---|
| 0.3967 | 0.8596 | 0.135276 | P4/nmm:2 |
| a(nm) | c(nm) | V (nm$^3$) | Space group |
| 0.3967 | 0.8596 | 0.135276 | P4/nmm:2 |
| S | $R_{wp}$ (%) | $R_b$ (%) | $R_{exp}$ (%) |
| 1.162 | 4.553 | 3.480 | 3.918 |

Before refinement

| | x/a | y/b | z/c | Occupancy |
|---|---|---|---|---|
| $Nd^{3+}$ | 0.25 | 0.25 | 0.1394 | 1 |
| $Fe^{2+}$ | 0.75 | 0.25 | 0.5 | 1 |
| As | 0.25 | 0.25 | 0.6579 | 1 |
| $O^{2-}$ | 0.75 | 0.25 | 0.0 | 0.8 |
| $F^{1-}$ | 0.75 | 0.25 | 0.0 | 0.2 |

After refinement

| | x/a | y/b | z/c | Occupancy |
|---|---|---|---|---|
| $Nd^{3+}$ | 0.25 | 0.25 | 0.1392 | 1 |
| $Fe^{2+}$ | 0.75 | 0.25 | 0.5 | 1 |
| As | 0.25 | 0.25 | 0.6580 | 0.96 |
| $O^{2-}$ | 0.75 | 0.25 | 0.0 | 0.8 |
| $F^{1-}$ | 0.75 | 0.25 | 0.0 | 0.2 |

**Sample: Nd-P0.05**

| a(nm) | c(nm) | V (nm$^3$) | Space group |
|---|---|---|---|
| 0.3968 | 0.8582 | 0.133635 | P4/nmm:2 |
| a(nm) | c(nm) | V (nm$^3$) | Space group |
| 0.3968 | 0.8582 | 0.133635 | P4/nmm:2 |
| S | $R_{wp}$ (%) | $R_b$ (%) | $R_{exp}$ (%) |
| 1.476 | 5.041 | 3.748 | 3.419 |

Before refinement

| | x/a | y/b | z/c | Occupancy |
|---|---|---|---|---|
| $Nd^{3+}$ | 0.25 | 0.25 | 0.1404 | 1 |
| $Fe^{2+}$ | 0.75 | 0.25 | 0.5 | 1 |
| As | 0.25 | 0.25 | 0.6541 | 0.95 |
| P | 0.25 | 0.25 | 0.6541 | 0.05 |
| $O^{2-}$ | 0.75 | 0.25 | 0.0 | 0.8 |
| $F^{1-}$ | 0.75 | 0.25 | 0.0 | 0.2 |

After refinement

| | x/a | y/b | z/c | Occupancy |
|---|---|---|---|---|
| $Nd^{3+}$ | 0.25 | 0.25 | 0.1403 | 1 |
| $Fe^{2+}$ | 0.75 | 0.25 | 0.5 | 1 |
| As | 0.25 | 0.25 | 0.6542 | 0.92 |
| P | 0.25 | 0.25 | 0.6542 | 0.04 |
| $O^{2-}$ | 0.75 | 0.25 | 0.0 | 0.8 |
| $F^{1-}$ | 0.75 | 0.25 | 0.0 | 0.2 |

**Sample: Nd-Sb0.05**

| a(nm) | c(nm) | V (nm$^3$) | Space group |
|---|---|---|---|
| 0.3962 | 0.8560 | 0.134370 | P4/nmm:2 |
| a(nm) | c(nm) | V (nm$^3$) | Space group |
| 0.3962 | 0.8560 | 0.134370 | P4/nmm:2 |
| S | $R_{wp}$ (%) | $R_b$ (%) | $R_{exp}$ (%) |
| 1.513 | 5.402 | 3.984 | 3.569 |

Before refinement

| | x/a | y/b | z/c | Occupancy |
|---|---|---|---|---|
| $Nd^{3+}$ | 0.25 | 0.25 | 0.1462 | 1 |
| $Fe^{2+}$ | 0.75 | 0.25 | 0.5 | 1 |
| As | 0.25 | 0.25 | 0.6556 | 0.95 |
| Sb | 0.25 | 0.25 | 0.6556 | 0.05 |
| $O^{2-}$ | 0.75 | 0.25 | 0.0 | 0.8 |
| $F^{1-}$ | 0.75 | 0.25 | 0.0 | 0.2 |

After refinement

| | x/a | y/b | z/c | Occupancy |
|---|---|---|---|---|
| $Nd^{3+}$ | 0.25 | 0.25 | 0.1456 | 1 |
| $Fe^{2+}$ | 0.75 | 0.25 | 0.5 | 1 |
| As | 0.25 | 0.25 | 0.6563 | 0.86 |
| Sb | 0.25 | 0.25 | 0.6563 | 0.04 |
| $O^{2-}$ | 0.75 | 0.25 | 0.0 | 0.8 |
| $F^{1-}$ | 0.75 | 0.25 | 0.0 | 0.2 |

Nd-(Sb+3)0.05

| a(nm) | c(nm) | V (nm $^3$) | Space group |
|---|---|---|---|
| 0.3962 | 0.8560 | 0.134370 | P4/nmm:2 |
| a(nm) | c(nm) | V (nm $^3$) | Space group |
| 0.3962 | 0.8560 | 0.134370 | P4/nmm:2 |
| S | $R_{wp}$ (%) | $R_b$ (%) | $R_{exp}$ (%) |
| 1.513 | 5.402 | 3.985 | 3.570 |

| | x/a | y/b | z/c | Occupancy |
|---|---|---|---|---|
| $Nd^{3+}$ | 0.25 | 0.25 | 0.1462 | 1 |
| $Fe^{2+}$ | 0.75 | 0.25 | 0.5 | 1 |
| As | 0.25 | 0.25 | 0.6556 | 0.95 |
| $Sb^{3+}$ | 0.25 | 0.25 | 0.6556 | 0.05 |
| $O^{2-}$ | 0.75 | 0.25 | 0.0 | 0.8 |
| $F^{1-}$ | 0.75 | 0.25 | 0.0 | 0.2 |

| | x/a | y/b | z/c | Occupancy |
|---|---|---|---|---|
| $Nd^{3+}$ | 0.25 | 0.25 | 0.1457 | 1 |
| $Fe^{2+}$ | 0.75 | 0.25 | 0.5 | 1 |
| As | 0.25 | 0.25 | 0.6563 | 0.88 |
| $Sb^{3+}$ | 0.25 | 0.25 | 0.6563 | 0.03 |
| $O^{2-}$ | 0.75 | 0.25 | 0.0 | 0.8 |
| $F^{1-}$ | 0.75 | 0.25 | 0.0 | 0.2 |

Nd-(Sb+5)0.05

| a(nm) | c(nm) | V (nm $^3$) | Space group |
|---|---|---|---|
| 0.3962 | 0.8560 | 0.134370 | P4/nmm:2 |
| a(nm) | c(nm) | V (nm $^3$) | Space group |
| 0.3962 | 0.8560 | 0.134370 | P4/nmm:2 |
| S | $R_{wp}$ (%) | $R_b$ (%) | $R_{exp}$ (%) |
| 1.513 | 5.401 | 3.985 | 3.570 |

| | x/a | y/b | z/c | Occupancy |
|---|---|---|---|---|
| $Nd^{3+}$ | 0.25 | 0.25 | 0.1462 | 1 |
| $Fe^{2+}$ | 0.75 | 0.25 | 0.5 | 1 |
| As | 0.25 | 0.25 | 0.6556 | 0.95 |
| $Sb^{5+}$ | 0.25 | 0.25 | 0.6556 | 0.05 |
| $O^{2-}$ | 0.75 | 0.25 | 0.0 | 0.8 |
| $F^{1-}$ | 0.75 | 0.25 | 0.0 | 0.2 |

| | x/a | y/b | z/c | Occupancy |
|---|---|---|---|---|
| $Nd^{3+}$ | 0.25 | 0.25 | 0.1457 | 1 |
| $Fe^{2+}$ | 0.75 | 0.25 | 0.5 | 1 |
| As | 0.25 | 0.25 | 0.6563 | 0.88 |
| $Sb^{5+}$ | 0.25 | 0.25 | 0.6563 | 0.03 |
| $O^{2-}$ | 0.75 | 0.25 | 0.0 | 0.8 |
| $F^{1-}$ | 0.75 | 0.25 | 0.0 | 0.2 |

Table 2. The bond length, bond angle and thickness of layer of $NdFeAsO_{0.8}F_{0.2}$unit cell

| sample | | Nd-P 0.05 | Nd-1111 | Nd-Sb 0.05 | Nd-$Sb^{+3}$0.05, Nd-$Sb^{+5}$ 0.05 |
|---|---|---|---|---|---|
| Bond angle | α: As-Fe-As (deg.) | 112.59(11) | 111.20(11) | 111.93(11) | 111.93(11) |
| | β: As-Fe-As (deg.) | 107.93(6) | 108.61(6) | 108.26(6) | 108.26(6) |
| | η: (O, F)- Nd- (O, F) (deg.) | 117.5(2) | 117.8(2) | 115.6(2) | 115.6(2) |
| Bond length | t: Fe-As (nm) | 0.2385(3) | 0.2404(3) | 0.2390(3) | 0.2390(3) |
| | y: Nd-O (nm) | 0.2321(4) | 0.2316(4) | 0.2340(4) | 0.2341(4) |
| | 2d: thickness of Fe-As layer (nm) | 0.2646(7) | 0.2716(3) | 0.2675(8) | 0.2675(8) |
| | 2h: thickness of Nd-O layer(nm) | 0.2408(1) | 0.2393(1) | 0.2492(7) | 0.0145(7) |
| | F: distance of Fe-As and Nd-O layers(nm) | 0.1763(6) | 0.1743(2) | 0.1695(7) | 0.1694(8) |

As you know that the integral breadth of the diffraction peaks can be linked to the apparent size of the crystals and to their microstrains. Williamson and Hall method has suggested a simple method for solving this problem. It works by considering that both the limited size of the crystals and the presence of crystallographic distortions lead to Lorentzian intensity distributions. If we denote by $\beta_p$ the pure breadth and by $\beta^T$ and $\beta^D$ the breadths related to size and microstrains, respectively, then we obtain:

$$\beta_P = \beta^T + \beta^D \quad (1)$$

and therefore:

$$\beta cos\theta/\lambda = K/D + \eta\, sin\theta/\lambda \quad (2)$$

where K is Scherrer's constant, β and θ are full width at half maximum and diffraction degree for each peak, respectively [52]. If we plot (β cos θ)/λ according to (sin θ)/λ, we get a straight line with a y-intercept equal to the inverse of the size and a slope equal to the value of the microstrains $\eta$ (See **Fig. 4**).

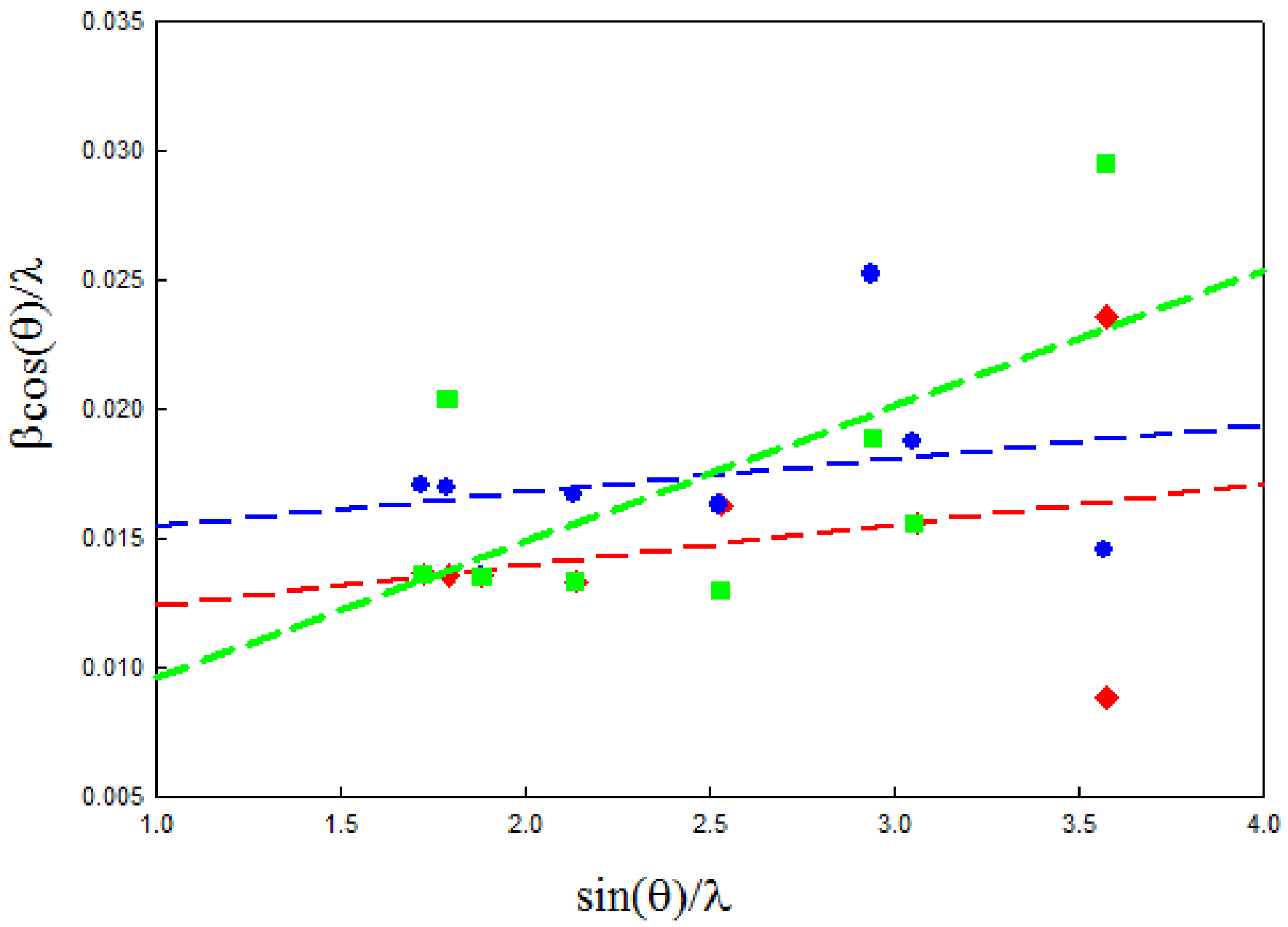

*Figure 4. Williamson–Hall plot of Nd-1111(blue diamond dots), Nd-Sb0.05(red square dots) and Nd-P0.05(green circle dots)*

The results of this computation for three samples are given in **table 3**.

*Table 3. Williamson–Hall equation, microstrain and crystalline size for all samples*

| Sample | Williamson-Hall equation | Microstrain η (%) | Crystalline size D (nm) |
|---|---|---|---|
| Nd-P0.05 | y= 0.0053x+0.0043 | 0.53 | 232.55 |
| Nd-1111 | y=0.0043x+0.0079 | 0.43 | 126.58 |
| Nd-Sb0.05 | y= 0.0046x+0.0047 | 0.46 | 212.77 |

According to the results of **table 3** and **table 2**, microstrain η is equal to 0.43% for the Nd-1111 sample which the volume of its unit cell is 0.135276 $nm^3$. For the Nd-P0.05 sample, since the phosphorus atom has an electron layer less than arsenic atom, also the volume of its unit cell is 0.133635 $nm^3$, therefore, the lattice becomes shrinking and its microstrain increasing to 0.53%. For Nd-Sb0.05 sample, the antimony atom has an electron layer more than arsenic and also the volume of its unit cell is 0.134370 $nm^3$, so again, the lattice will shrink fewer than the unit cell of Nd-P0.05 and it results in fewer microstrain than this sample.

As be shown in **table 3**, the crystalline size of samples increases for Nd-Sb0.05 and Nd-P0.05 respect to Nd-1111 sample, respectively. The greater microstrain in the Nd-P0.05, can be due to the smaller size of the phosphorus atom than Antimony and Arsenic atoms. This causes to more distortion of the unit cell.

We used the 4-probe contact method for obtaining superconductivity parameters of our samples. The 20 K Closed Cycle Cryostat (QCS 101), ZSP Croygenics Technology (Iran), was used for measurement of superconductivity behavior of samples. The normalized resistivity as a function of temperature is shown in **Fig. 5**. As be showed in **table 4**, for Nd-1111sample, the electrical resistivity gradually decreases by cooling and the onset of the resistive transition is 62 K, and finally, it reaches to zero at $T_{offset}$=50 K. The critical temperature was obtained at 56 K for this sample[14]. This sample exhibits as a metallic behavior in normal state. It was observed that by substitution of P in Nd-1111, the superconductivity state eliminates and the structural transition temperature appears at 140 K.

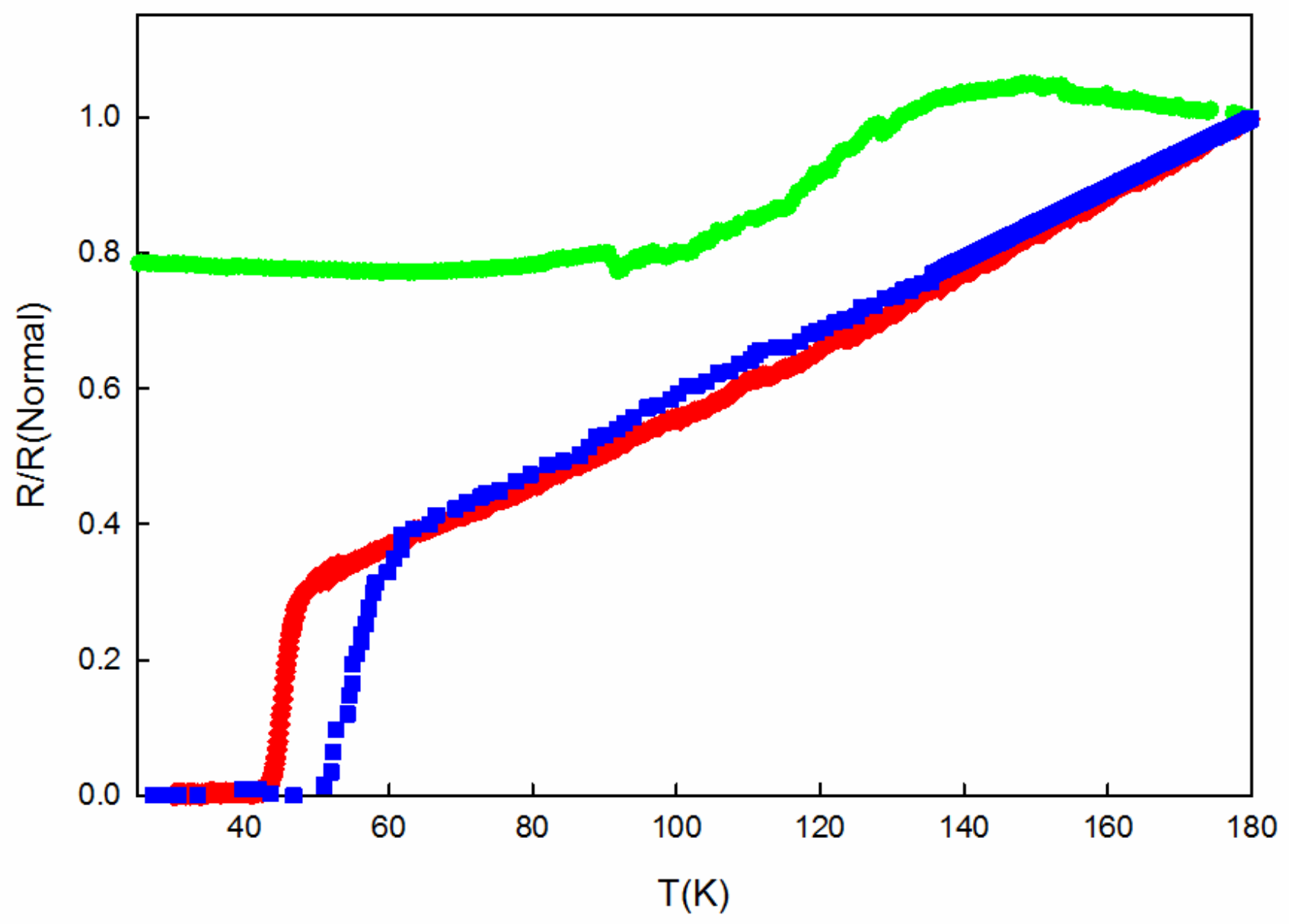

*Figure 5. The normalized resistivity of Nd-1111(square blue dots), Nd-P0.05 (circle green dots) and Nd-Sb0.05 (diamond red dots)*

*Table 4. superconductivity temperatures of the Nd-1111, Nd-P0.05, Nd-Sb0.05 samples*

| sample | $T_{onset}$(K) | $T_{offset}$(K) | $T_C$(K) | ΔT(K) |
|---|---|---|---|---|
| Nd-1111 | 62 | 50 | 56 | 12 |
| Nd-Sb0.05 | 48 | 44 | 46 | 4 |
| Nd-P0.05 | - | - | - | - |

At a higher temperature than the structural transition temperature, the behavior of the Nd-P0.05 sample is similar to the insulator. It can be due to the phosphorus atom. It is the only element in the fifteenth group of periodic tables that has insulating behavior. For Nd-Sb0.05 sample, the electrical resistivity decreases by cooling and then rapidly drop at $T_{onset}$=48 K, and finally, it reaches to zero at $T_{offset}$=44 K. The critical temperature was obtained at 46 K for this sample.

The shape of the $FeAs_4$ tetrahedron seems to play a conclusive role with, in particular, the Fe-As bond length having a robust effect on the transition temperature superconductivity [35,53,20,24,22,54,12]. The bond length of Fe-As in the Nd-1111 sample is 0.2404 nm and it decreases by substitution of P/As and Sb/As. It was observed that the bond length of Fe-As, the β bond angle in this structure and the thickness of Fe-As layer, increase by the increasing the transition temperature superconductivity. Our work confirms other research works for this claim [55,24,22].

The surface morphology and the grain connectivity of the all samples have been determined using Scanning Electron Microscopy (SEM). These three samples imaged with the TESCAN VEGA3 LMU with magnification 60.000e3. The SEM images illustrated in **Figs. 6-(a-d)** for the Nd-1111, Nd-Sb0.05 and Nd-P0.05.

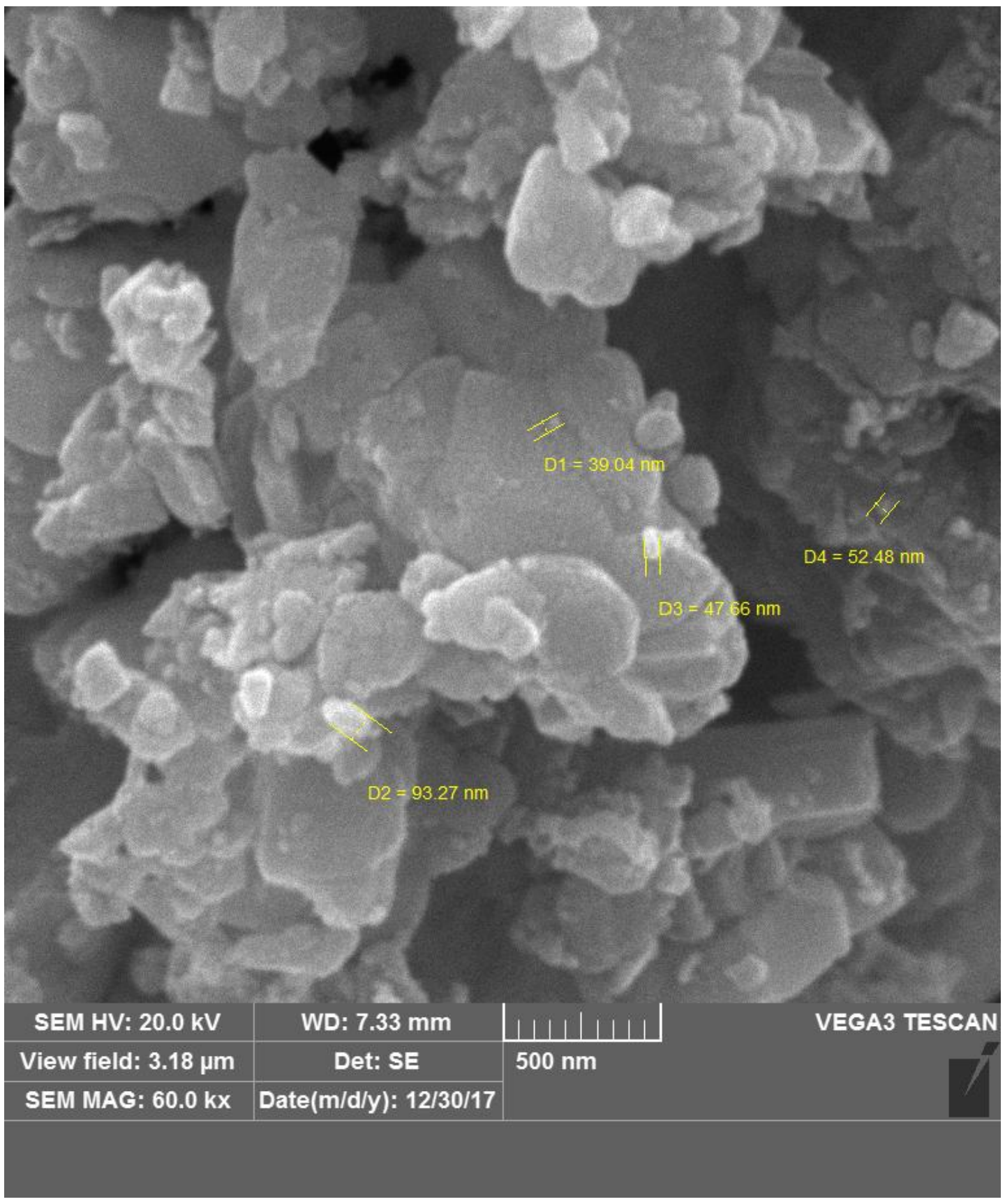

*Figure 6- a. SEM image for the Nd-1111*

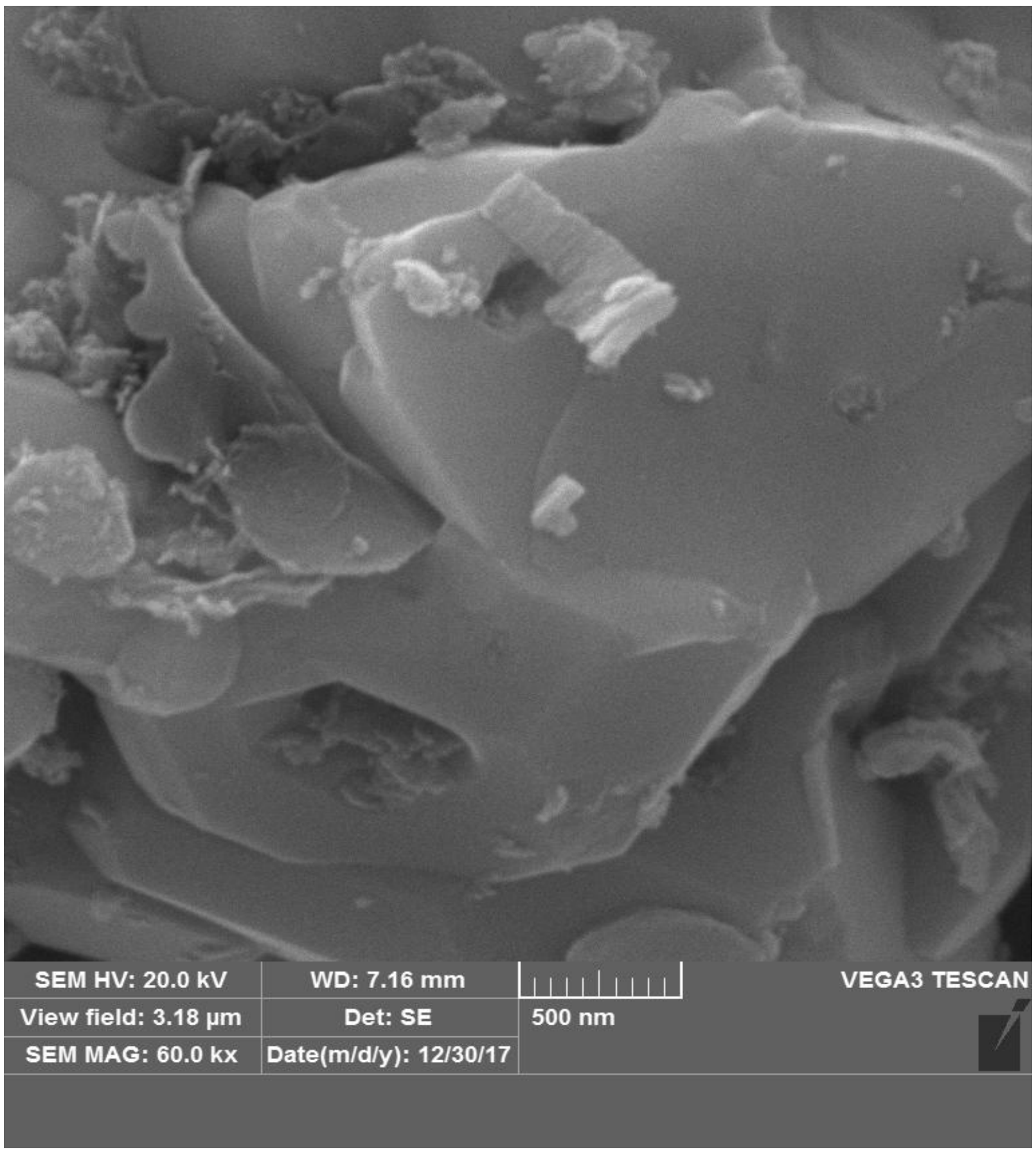

*Figure 6- b. SEM image for the Nd-Sb0.05*

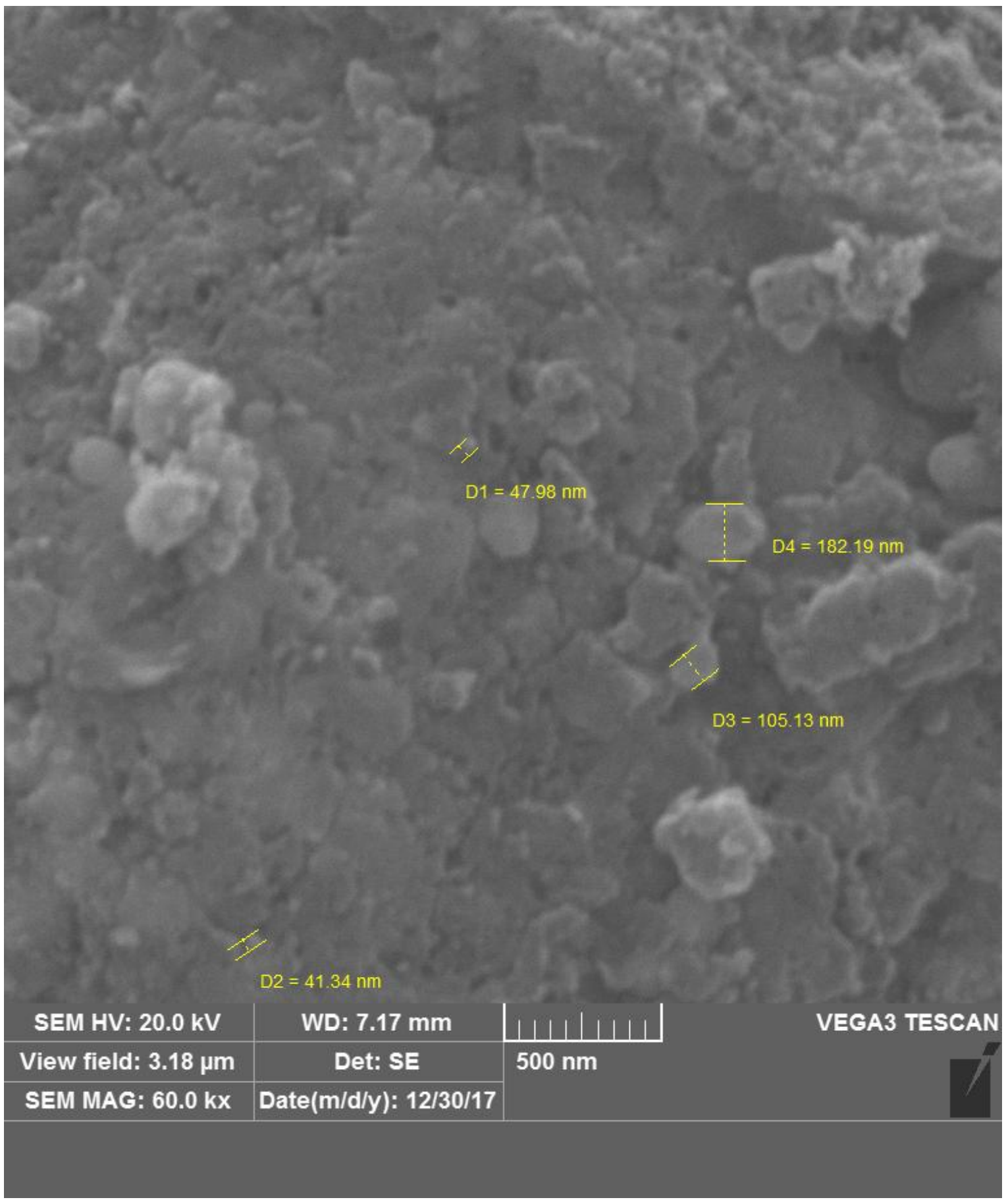

*Figure 6- c. SEM image for the Nd-Sb0.05*

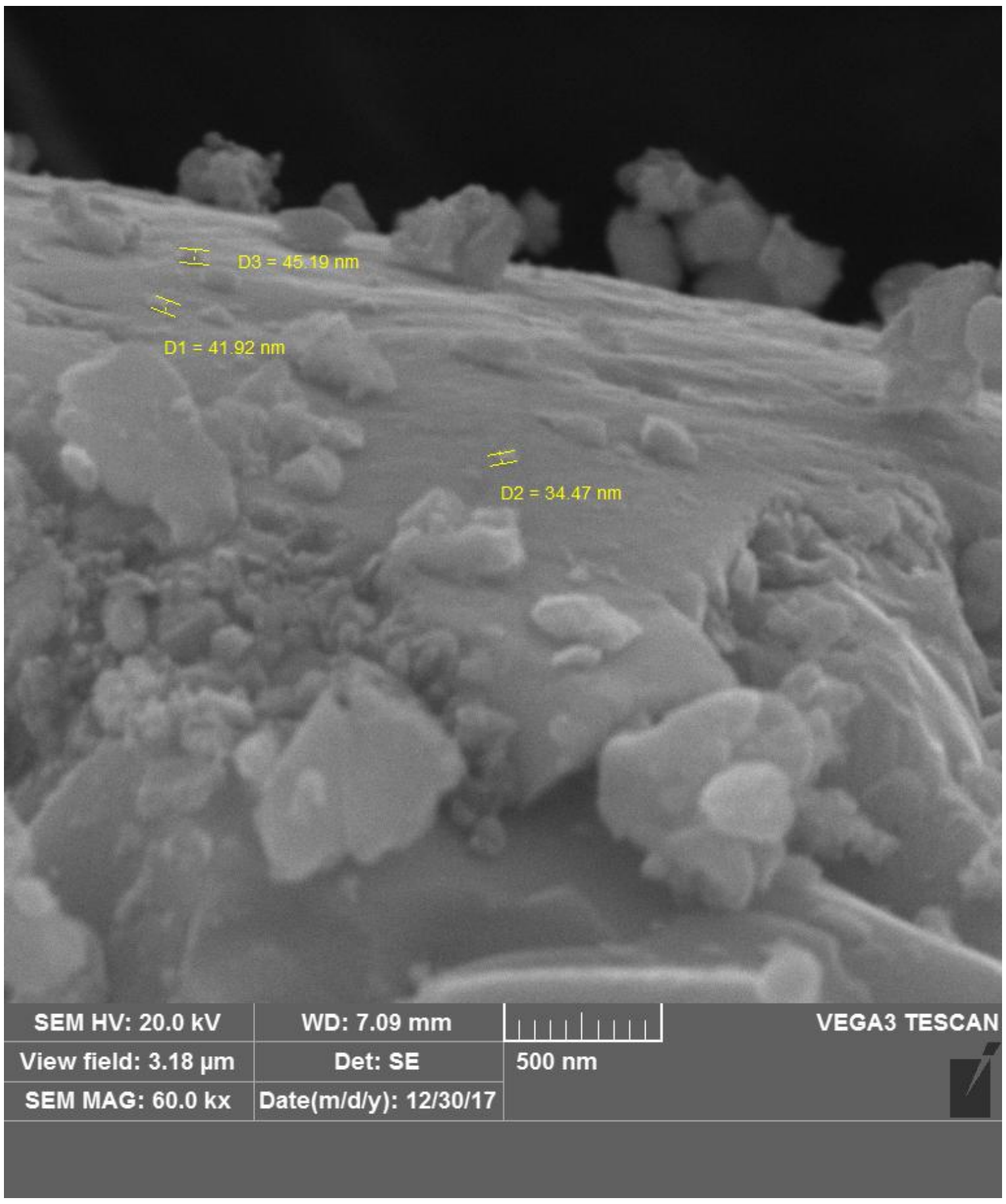

*Figure 6- d. SEM image for the Nd-P0.05*

The SEM images clearly represent the crystalline structure of sample. Since the edges of the crystal grains are sharp (See **Figs 6-(a-d)**), the temperatures selected in the manufacturing method are well chosen and implying the correct preparation method.

As be shown in **Fig. 6-a** for Nd-1111, **Fig. 6-b** and **Fig.6-c** for Nd-Sb0.05 and **Fig. 6-d** for Nd-P0.05, the surface morphology of these samples is the mixture of flake-type and nodule-type grains and the grain are distributed randomly with the grains size between 39nm and more than 100 nm, 40nm and more than 300 nm and 34nm and more than 250 nm, respectively.

The apparent shape of grain for all samples can also be confirmed by the sample manufacturing method.

## 4. Conclusion

In summary, based on Rietveld refinement of XRD patterns of samples, it was showed that the all samples have similar structural properties as tetragonal crystal structure with p4/nmm:2 space group with different unit cell parameters. The “a” unit cell parameter of samples increasing by substitution of P/As to Sb/As. The “c” unit cell parameter of samples due to the electronegativity of isovalent atoms, increases from Nd-P0.05, Nd-Sb0.05 and Nd-1111, respectively. Our results show that the Arsenic isovalent doping increased the α angle and decreased the β angle, bond length of the Fe-As and the thickness of Fe-As layer. The substitution of P/As in nominal composition of Nd-1111 suppressed the superconductivity. We have observed a structural phase transition at 140 K in Nd-P0.05. Also, the $T_C$ had decreased from 56 K to 46K in Nd-Sb0.05 by substitution of Sb/As in nominal composition of Nd-1111. The microstrain of Nd-1111 was increased by substitution of P/As more than Sb/As. It seems that the lattice shrinkage can be related with the $T_C$ and it becomes maximum when the FeAs4 tetrahedron of lattices close to the regular tetrahedron.

## Acknowledgements

The authors would like to thank the Prof. M. Akhavan for the measurement set up laboratory and they acknowledge the Vice Chancellor Research and Technology of Alzahra university.